\documentclass[12pt]{article}
\usepackage{amssymb}
\begin{document}
\renewcommand \baselinestretch{1.3}
\renewcommand{\thesection}{\arabic{section}.}
\textwidth 15.5cm \oddsidemargin 0.75cm \evensidemargin 0.75cm
\topmargin -0.8cm \textheight 20.8cm
\newcommand{\BE}{\begin{equation}}
\newcommand{\EE}{\end{equation}}
\newcommand{\half}{{\scriptstyle{\frac{1}{2}}}}
\newcommand{\boldsymbol}{\bf}
%


\begin{titlepage}

\vspace*{1mm}
\begin{center}

{\LARGE{\bf  Two mass scales for the Higgs field? }}

\end{center}

\vspace*{0.1cm}

\begin{center}
{\large
Paolo Cea$^{\dagger}$\protect\footnote{Electronic address:
paolo.cea@ba.infn.it},
Maurizio Consoli$^{\ddagger}$\protect\footnote{Electronic address:
maurizio.consoli@ct.infn.it}
and Leonardo Cosmai$^{\dagger}$\protect\footnote{Electronic address:
leonardo.cosmai@ba.infn.it} \\[0.5cm]
$^{\dagger}${\em INFN - Sezione di Bari, I-70126 Bari, Italy} \\[0.3cm]
$^{\ddagger}${\em INFN - Sezione di Catania,  I-95129 Catania, Italy } }
\end{center}

\begin{center}
{\bf Abstract}
\end{center}

\par\noindent In the original version
of the theory, the driving mechanism for spontaneous symmetry
breaking  was identified in the pure scalar sector. However, this
old idea requires a heavy Higgs particle that, after the discovery
of the 125 GeV resonance, seems to be ruled out. We argue that this
is not necessarily true. If the phase transition is weakly first
order, as indicated by most recent lattice simulations, one should
consider those approximation schemes that are in agreement with this
scenario. Then, even in a simple one-component theory, it becomes
natural to introduce two mass scales, say $M_h$ and $m_h$ with $m_h
\ll M_h$. This resembles the coexistence of phonons and rotons in
superfluid helium-4, which is the non-relativistic analogue of the
scalar condensate, and is potentially relevant for the Standard
Model. In fact, vacuum stability would depend on $M_h$ and not on
$m_h$ and be nearly insensitive to the other parameters of the
theory (e.g. the top quark mass). By identifying $m_h=125$ GeV, and with 
 our previous estimate from lattice simulations $M_h= 754 \pm 20 ~\rm{(stat)}  \pm 20 ~\rm{(syst)}$   GeV, 
we thus get in touch with a recent, independent analysis of the ATLAS + CMS data which claims
experimental evidence for a scalar resonance around $700$ GeV.

\end{titlepage}

\section{Introduction}

The phenomenon of spontaneous symmetry breaking, that is the
generation of all particle masses from the vacuum expectation value
$\langle \Phi \rangle \neq 0$ of the Higgs field, is an essential
ingredient of the Standard Model. The idea is remarkably simple and
has a long history which dates back to more than fifty years ago
\cite{higgs,englert}. Moreover, there has been an important
experimental confirmation after the observation, at the Large Hadron
Collider of CERN \cite{LHC}, of a narrow scalar resonance, of mass
$m_h \sim 125 $ GeV whose phenomenology fits well with the
perturbative predictions of the theory. Thus, one might think that,
by now, very little remains to be understood.

Yet, a notable aspect of the theory remains unclear, namely the
order of the phase transition in pure $\Phi^4$ theories. As we will
illustrate, this is an important issue that may have substantial
phenomenological implications. In this respect, recent lattice
simulations of $\Phi^4$ theory in four space-time dimensions
\cite{lundow,japan} have added new precious evidence. In fact, these
calculations, performed in the Ising limit of the theory with
different algorithms, indicate that on the largest lattices
available so far the phase transition is (weakly) first-order.

With this non-perturbative numerical evidence, to explore the
possible implications, it would be natural to restrict to those
analytical approximations that indeed predict a weakly first-order
scenario. However, since there are several subtleties, for sake of
clarity we will first re-capitulate the general problem along the
lines of Refs.~\cite{mech,stevenson2009}.

Let us therefore start from scratch with the classical $\Phi^4$
potential \BE V_{\rm class}=\frac{1}{2} r_B \Phi^2+ \frac{1}{4!}
\lambda\Phi^4 \EE which gives an unambiguous indication: as one
varies the bare $r_B$ mass parameter, there is a second-order phase
transition at $r_B = 0$.

In the quantum theory, the question is more subtle and, to be
formulated, requires to consider the mass squared parameter, say
$m^2_\Phi$, introduced by quantizing the theory in the symmetric
phase at $\langle \Phi \rangle=0$. Clearly, this symmetric vacuum is
{\it locally} stable if its excitations have a physical mass
$m^2_\Phi> 0$. However, is this vacuum also {\it globally} stable?
Namely, could the phase transition actually be first order, i.e.
occurring at some very small but still positive $m^2_\Phi$ as
originally suggested by Coleman and Weinberg \cite{CW}?

Here, for a pure $\Phi^4$ (no gauge couplings), the standard
approximation methods for the quantum effective potential $V_{\rm
eff}(\phi)$ give contradictory results \cite{CW}. The
straightforward one-loop approximation predicts a first-order
transition occurring at a small critical value of mass squared,
$m^2_\Phi = m^2_c  > 0$. On the other hand, the usual
Renormalization Group (RG) ``improvement'', obtained by resumming
the leading-logarithmic terms, predicts a second-order transition at
$m^2_\Phi = 0$. The conventional view is that the latter result is
trustworthy while the former is not. The argument is that, for
$0\leq m^2_\Phi < m^2_c$ , the one-loop potential's non-trivial
minimum occurs where the one-loop ``correction'' term is as large as
the tree-level term. However, also this standard RG-improved result
can hardly be trusted because amounts to re-summing a geometric
series of leading logs that is actually a divergent series
\cite{mpla1996}.

To understand the reason of the discrepancy, a crucial observation
is that the quanta of the symmetric phase, the ``phions''
\cite{mech}, besides the +$\lambda \delta^{3}(\bf r)$ contact
repulsion, also feel a $-\lambda^2 \frac {e^{-2 m_\Phi r}}{r^3}$
attraction which shows up at the one-loop level and whose range
becomes longer and longer in the $m_\Phi \to 0$ limit. By taking
into account both effects, a calculation of the energy density in
the dilute-gas approximation \cite{mech}, which is equivalent to the
one-loop potential, indicate that for small $m_\Phi$ the
lowest-energy state is not the empty state with no phions but a
state with a non-zero density of phions Bose condensed in the
zero-momentum mode. The instability corresponds to spontaneous
symmetry breaking and happens when the phion's physical mass
$m^2_\Phi$ is still positive; it does not wait until $m^2_\Phi$
passes through zero and becomes negative. Though the critical mass
$m^2_c$ is extremely small so that it is a very weak first-order
transition which becomes indistinguishable from a second-order
transition if one does not look on a fine enough scale.

Now, since symmetry breaking originates from two qualitatively
different competing effects, one can understand why the standard
RG-analysis fails. In fact, the one-loop attractive term is {\it
ultraviolet finite}. Therefore, the correct way to include higher
order terms is to renormalize {\it both} the tree-level repulsion
and the long-range attraction, as in a theory with {\it two}
coupling constants. This strategy, which is clearly different from
the usual one, has been implemented by Stevenson
\cite{stevenson2009}. By avoiding double counting, he has shown that
one-loop result and its RG-group improvement, in this new scheme,
now agree very well so that the weak first-order scenario is
confirmed.

On the other hand, as an additional check, one can also compare with
other non-perturbative approximations, for instance the Gaussian
approximation \cite{gaussian} that, in principle, should be the most
natural scheme. In fact, at least in the continuum limit, it
respects the generally accepted ``triviality'' of the theory in 3+1
dimensions. This other calculation produces a result in agreement
with the one-loop effective potential \cite{zeit}. The agreement is
not because it contains no non-vanishing corrections beyond the
one-loop level; it does but those additional terms do not alter the
functional form of the result. Once more, the weak first-order
scenario in $\Phi^4$ theories is confirmed.

Notwithstanding, all this has gone practically unnoticed within the
high-energy community. The main reason dates back again to Coleman
and Weinberg \cite{CW} who observed that no conflict, between
one-loop potential and its standard RG-improvement, arises in the
presence of gauge bosons, for instance in scalar electrodynamics, at
least if the scalar self-coupling is not too large. Because of this
result, which is considered as the only relevant for the Standard
Model, the problem and the implications of the phase transition in
pure $\Phi^4$ theories have been left aside.

However, once all couplings are put on the same level, the scalar
sector becomes strongly constrained. Therefore, the original picture
where symmetry breaking was only determined by the pure scalar
sector, the other couplings just producing small corrections, has
been abandoned. The consistency of that original picture would, in
fact, require a substantially heavy Higgs boson which by now seems
to be in conflict with experiment.

Our scope in this Letter is to show that this is not necessarily
true. If some aspects of the phase transition in $\Phi^4$ theories
have been overlooked there may be some ambiguity concerning the role
and the meaning of what, in this context, is understood by ``Higgs
particle mass''. To this end, we will re-reconsider in Sect.2 the
one-loop calculation of the effective potential (or the equivalent
Gaussian approximation) in the cutoff theory. Formally, there is
nothing new in this elementary calculation. But, if this were
accepted as the correct description of symmetry breaking, its {\it
interpretation} could now become completely different. Namely there
might be {\it two} vastly different mass scales in the broken phase,
say $M_h$ and $m_h$ with $m_h \ll M_h$. The important point is that
the stability of the vacuum depends on the larger $M_h$ and not on
$m_h$. Therefore, spontaneous symmetry breaking could be determined
by the pure scalar sector regardless of the other parameters of the
theory (e.g. the vector boson and top quark mass).

To help physical intuition, one can exploit the analogy  with the
non-relativistic limit of the scalar condensate, namely superfluid
helium-4. The elementary constituents of the superfluid are the
helium-4 atoms but at low energy only collective excitations of the
system are observable, first its gapless compressional modes (the
phonons) and then the vortical modes (rotons) that possess an energy
gap. For very low momenta $ {\bf k} \to 0$ only phonons propagate.
But, by increasing the energy also rotons can be excited. In this
analogy the lower mass $m_h$ would correspond to phonons while the
heavy mass $M_h$ would play the role of mass gap for the roton
branch.

Then, with our previous estimate \cite{cea2003,cea2012} from lattice
simulations $M_h= 754 \pm 20 ~\rm{(stat)}  \pm 20 ~\rm{(syst)} $ GeV,  
we will get in touch with a
recent, independent analysis \cite{cea2019} of the ATLAS + CMS data
which claims evidence for a scalar resonance around $700$ GeV. These
more phenomenological aspects will be addressed in Sect.3.

\section{Two mass scales for the Higgs field?}

Let us assume the scalar $\Phi^4$ Lagrangian \BE {\cal L}= \half
(\frac{\partial \Phi}{\partial x_\mu})^2 -{{r_B \Phi^2}\over{2
}}-{{\lambda\Phi^4}\over{4! }} \EE and shift $\Phi= \phi + h(x)$.
The long discussion given in the Introduction indicates that the
one-loop potential, or the equivalent gaussian effective potential,
are expected to give the correct description of symmetry breaking.
Let us thus consider the self energy in the one-loop approximation
\BE \label{self} \Pi^{\rm 1-loop}(p)= - r_B - {{\lambda
\phi^2}\over{2 }} - {{\lambda}\over{2 }} A_0(M) +
{{\lambda^2\phi^2}\over{2 }}B_0(p,M,M) \EE where \BE
A_0(M)=\int{{d^4k}\over{(2\pi)^4 }}{{1}\over{k^2 + M^2 }} \EE and
\BE B_0(p,M,M)=\int{{d^4k}\over{(2\pi)^4 }}{{1}\over{[(p+k)^2 + M^2]
(k^2 + M^2)}} \EE Now, by fixing the mass counterterm  as in
Coleman-Weinberg, i.e. \BE r_B=-{{\lambda}\over{2 }} A_0(M=0)\EE one
finds their expression for the effective potential in the presence
of a large ultraviolet cutoff $\Lambda_s$ for the scalar sector
 \BE \label{veffeq}
V_{\rm eff}(\phi) =  \frac{\lambda}{4!} \phi^4 +\frac{\lambda^2}{256
\pi^2} \phi^4 \left[ \ln (\half \lambda \phi^2 /\Lambda^2_s ) -
\frac{1}{2} \right]  \EE whose first few derivatives are \BE
\label{vprime}
V'_{\rm eff}(\phi) =  \frac{\lambda}{6} \phi^3 +\frac{\lambda^2}{64
\pi^2} \phi^3 \ln (\half \lambda \phi^2 /\Lambda^2_s )  \EE and \BE
\label{vsecond}
V''_{\rm eff}(\phi) =  \frac{\lambda}{2} \phi^2
+\frac{3\lambda^2}{64 \pi^2} \phi^2 \ln (\half \lambda \phi^2
/\Lambda^2_s ) +\frac{\lambda^2\phi^2}{32\pi^2}   \EE This second
derivative is equivalent to compute $-\Pi^{\rm 1-loop}(p=0)$ in
Eq.(\ref{self}) by replacing the tree-level mass
$M^2(\phi)={{\lambda \phi^2}\over{2 }}$ as the mass which runs in
the loops. In standard perturbation theory, this would be the first
step of an iterative procedure where one starts with a zeroth-order
mass, say $M^{\rm 0-loop}$, and replace in the loops of $\Pi^{\rm
1-loop}$. Then, by performing corresponding renormalization of the
coupling constant, one can define the mass at one-loop, say $M^{\rm
1-loop}$. In general, to order n, $M^{\rm 0-loop}$ should be
replaced in the diagrams with n loops, $M^{\rm 1-loop}$ in the
diagrams with (n-1) loops and so on. In this way, together with
coupling constant and wave function renormalization, one can extend
the analysis to any desired order.

However, by following this strategy one predicts the wrong
second-order phase transition. Instead, for the reasons explained in
the Introduction, we expect that it is the one-loop effective
potential to display the correct physical interpretation. At its
minima, say $\phi=\pm v$, and by defining $M^2(\pm v)= M^2_h$, this
gives two different informations: 
\vskip 5 pt ~~~i) its depth
$V_{\rm eff}(\pm v)\sim -M^4_h$

~~ii) its quadratic shape $V''_{\rm eff}(\pm v)\equiv m^2_h \ll
M^2_h $ \vskip 5 pt \noindent On this basis, we will argue that the
two mass scales $m_h$ and $M_h$ describe the propagator in two
vastly different regions of momenta, respectively $p \to 0$ and $p^2
>> m^2_h$.

To explore the $p\to 0$ limit, let us first look at the minima of
$V_{\rm eff}(\phi)$ where  \BE \label{basic} M^2_h={{\lambda
v^2}\over{2 }}= \Lambda^2_s \exp( -{{32 \pi^2 \lambda}\over{3 }})
\EE so that \BE -\Pi^{\rm 1-loop}(p=0)= V''_{\rm eff}(\pm v) =
\frac{\lambda^2v^2}{32\pi^2}= \frac{\lambda}{16\pi^2}M^2_h\equiv
m^2_h  \EE and for large $L\equiv \ln \frac{\Lambda_s}{M_h}$ \BE
\label{small} m^2_h =\frac{M^2_h}{3L} \ll M^2_h \EE Notice that the
energy density depends on $M_h$ and {\it not} on $m_h$, because
 \BE V_{\rm eff}(\pm v)= -\frac{M^4_h}{128 \pi^2 } \EE
therefore the critical temperature at which symmetry can be restored
is  $k_BT_c\sim M_h$. This means that the stability conditions of
the broken phase depends solely on the large scale $M_h$ and not on
the much smaller scale $m_h$ which determines the propagator for $p
\to 0$ \BE G^{-1}(p)=p^2 -\Pi(p)=p^2 + m^2_h - m^2_h \int^1_0 dx
\ln\frac{p^2x(1-x) +M^2_h}{M^2_h}
 \EE
One can thus approximate the vanishing of the inverse propagator as
\BE p^2+ m^2_h- m^2_h ~\frac{p^2}{6M^2_h}=0 \EE By defining
$\alpha\equiv \frac {m^2_h}{6M^2_h } \ll 1$, in Minkowski space this
is a pole whose approximate location is very close to $m^2_h$ \BE
(p^2_0 -{\bf p }^2)\sim \frac{m^2_h}{1-\alpha} \sim m^2_h(1+ \alpha)
\EE Let us now consider the higher$-p^2$ region. Strictly speaking,
the effective potential generates the vertices at $p=0$. However,
insight into $p\neq 0$ can be obtained by comparing with the general
expression of the zero-point energy: the trace of the log of the
inverse propagator $G^{-1}(p)= p^2- \Pi(p)$, namely \BE
\label{general} \frac{1}{2} \int {{ d^4 p}\over{(2\pi)^4}} \ln
(p^2-\Pi(p)). \EE After subtractions, its value can be reproduced by
imposing appropriate lower and upper limits to the $p$-integration
in the logarithmic divergent part
 \BE \label{connection}
-\frac{1}{4}\int {{ d^4 p}\over{(2\pi)^4}} \frac{\Pi^2(p)}{p^4} .
 \EE
so that, for $\phi$ close to $\pm v$, one can compare directly with the
one-loop form \BE
-\frac{M^4(\phi)}{64\pi^2}[\ln\frac{\Lambda^2_s}{M^2(\phi)}
+\frac{1}{2}] \EE It is then clear that $M^2(\phi)$ cannot be a
purely infrared scale, i.e. whose only role is to regulate the
infrared divergences. In fact, besides entering the log, it
controls, through the value of  $M^4(\phi)$, an effect which gets contributions from the whole range
of $p$. Therefore the corresponding value $M^2_h$ for $\phi= \pm v$
will reflect the magnitude of $|\Pi(p)|$ in some {\it intermediate}
region $m^2_h \ll p^2 \ll \Lambda^2_s$ (it cannot be $p \to 0$ since
we have seen that there $|\Pi(0)|=V''_{\rm eff}(\pm v)=m^2_h \ll
M^2_h$).

Note that we are not saying that $M^2_h$ is the higher-$p^2$ limit
of $\Pi^{\rm 1-loop}(p)$. Trusting in the one-loop potential, only
the two basic relations $V_{\rm eff}(\pm v)\sim -M^4_h$ and
$V''_{\rm eff}(\pm v) = |\Pi(0)|= m^2_h \ll M^2_h$ are reliable. But
then, from Eq.(\ref{connection}) it follows that there must be a
higher momentum region $p^2
>>m^2_h$ where $|\Pi(p)|\sim M^2_h$. Together with the complementary $p \to 0$
region where $|\Pi(0)|\sim m^2_h$, this means that, even in a
one-component theory, the shifted field of the broken phase (the
equivalent of the Standard Model Higgs boson) can hardly be
considered a simple massive field. Somehow, it materializes in two
different mass scales, $m_h$ and $M_h$, whose quadratic ratio is
suppressed by the inverse logarithm of the cutoff $\Lambda_s$.

In a more complete derivation, the inverse propagator should then
emerge as a suitable interpolation\footnote{Sometimes, one can guess
the right form of the spectrum in two different limits but, as in
the case of the phonon-roton spectrum in superfluid helium-4,
describing the detailed transition between the two regimes remains a
difficult task. In our case, numerical evidence for two different
mass scales was found in lattice simulation of the spontaneously
broken phase in the Ising limit \cite{further}. To this end, the
high-momentum region of the connected propagator was fitted to have
an estimate of $M^2_h \equiv m^2_{\rm latt}$. Analogously, the
inverse zero-momentum two-point function was computed through the
lattice susceptibility to have an estimate of
${{1}\over{m^2_h}}\equiv \chi_{\rm latt}$. Then, the product
$m^2_{\rm latt}\chi_{\rm latt}=(M_h/m_h)^2$ was computed and found
to increase, consistently with a logarithmic trend, in the continuum
limit. On the other hand, no evidence for such two-scale structure
was found in the symmetric phase. There, the value of $m^2_{\rm
latt}$, fitted from the high-momentum propagator, describes the data
remarkably well down to $p=0$. } between these two regimes, say \BE
G^{-1}(p) = (p^2 + M^2_h) f (p^2/m^2_h) \EE with $f(p)\sim
(m_h/M_h)^2$ in the $p \to 0$ limit and $f (p^2/m^2_h) \to 1$ for
momenta $p^2 >> m^2_h$ . At present, as a definite example, by
defining the $\Phi^4$ theory as a limit where, from the very
beginning, one starts with a hard-sphere repulsion + non-local
long-range attraction, a form for such interpolating function is
given in Stevenson's Eqs.(16)-(22) of Ref.~\cite{stevenson2009}.
Note that his Eq.(23) should be read as $G^{-1}(p)$ and that he
considers the continuum limit $(m_h/M_h)^2 \to 0$. Then $f(x)$
becomes a step function which is unity for any finite $p$ except for
a discontinuity at $p=0$ where $f=0$. Up to this discontinuity in
the zero-measure set $p=0$, one then re-discovers the usual trivial
continuum limit with {\it only one} free massive particle.

As anticipated, an equivalent description is found in the Gaussian
approximation where one re-sums all one-loop bubbles. This other
calculation can be cast in a form which is similar to
Eq.(\ref{veffeq}) with a simultaneous re-definition of the mass and
of the classical background: \BE \label{vgauss}
V^G_{\rm eff}(\phi) =  \frac{\hat\lambda\phi^4}{4!} +
\frac{\Omega^4(\phi) }{64 \pi^2} \left[ \ln (\Omega^2(\phi)
/\Lambda^2_s ) - \frac{1}{2} \right]  \EE with \BE \hat \lambda=
\frac{\lambda } {1 + \frac{\lambda}{16 \pi^2} \ln \frac {\Lambda_s}{
\Omega(\phi)}   }  \EE and  \BE   \Omega^2(\phi)    =
\frac{\hat\lambda\phi ^2} {2 } \EE This explains why the one-loop
potential can also admit a non-perturbative interpretation. It is
the prototype of all gaussian and post-gaussian
\cite{stancu,tedesco} calculations where the energy density is given
as a classical background + zero-point energy of a field with a
$\phi-$dependent mass.

\section{Getting in touch with phenomenology}

The large $m_h-M_h$ difference reflects an effective potential which
is extremely flat because reaching its depth $V_{\rm eff}(\pm v)\sim
-M^4_h$ will take a very large distance if $V_{\rm eff}$ is plotted
in units of the $\phi-$field with second derivative $V''_{\rm
eff}(\phi=\pm v)= m^2_h$. For this reason, in refs.~\cite{zeit} a
large re-scaling~\footnote{We emphasize that this is the re-scaling
of the {\it vacuum} field and, as such, is quite unrelated to the
more conventional definition $Z=Z_{\rm prop}= 1 + O(\lambda)$ which
enters the residue of the {\it shifted} field propagator. By
``triviality'', the latter is constrained to approach unity in the
continuum limit. To better understand the difference, it is useful
to regard symmetry breaking as a true condensation phenomenon
associated with the macroscopic occupation the same quantum state
${\bf k}=0$. Then $\phi$ is related to the condensate while the
shifted field is related to the modes at ${\bf k} \neq 0$ which are
not macroscopically populated.} of the vacuum field $Z=Z_\phi $ was
introduced through the relation \BE V''_{\rm eff}(\pm v) = m^2_h=
\frac{\lambda}{16\pi^2}M^2_h\equiv \frac{M^2_h}{Z_\phi}\EE In this
way, one can define a re-scaled field $\phi_R$, with $\phi^2=Z_\phi
\phi^2_R$, such that the quadratic shape of the effective potential,
in terms of $\phi_R$, now matches exactly with $M^2_h$.

Therefore, a question naturally arises: if symmetry breaking were
generated in the pure scalar sector, when one couples scalar and
gauge fields, which is the correct definition of the expectation
value $\langle \Phi \rangle\sim $ 246 GeV entering the W mass
$M^2_w\sim\frac{g^2 \langle \Phi \rangle^2}{4}$ (and then the Fermi
constant through $\frac{G_F}{\sqrt{2}}\sim \frac{g^2}{8 M^2_w}$)? In
fact, this $\langle \Phi \rangle$ could be the same $v$ considered
so far which in general, i.e. beyond the Coleman-Weinberg limit, is
related to $M_h$ through a relation similar to Eq.(\ref{basic})
($L\equiv \ln \frac{\Lambda_s}{M_h})$ say \BE \label{large} M^2_h =
\frac{c_1 v^2}{L} \EE where $c_1$ is some constant. Or, instead, it
could be the much smaller $v_R$ \BE\label{basicnew} v^2_R
=\frac{v^2}{Z_\phi}= v^2~\frac{m^2_h}{M^2_h}= \frac{c_2 v^2}{L} \ll
v^2 \EE $c_2$ being another constant which replaces Eq.(\ref{small})
in the general case.

Now, in Ref.~\cite{zeit}, one argued as follows. $M_h$ determines
the vacuum energy, and thus the temperature $T_c$ of the phase
transition. In this sense, it is the natural cutoff-independent
quantity. At the same time, $\langle \Phi \rangle \sim $ 246 GeV is
a basic entry of the theory (as the electron mass and fine structure
constant in QED). Therefore, it would be natural to consider the
definition $v_R\equiv \langle \Phi \rangle\sim 246 $ GeV which is
finitely related to $M_h$ through some proportionality constant $K$
 \BE M_h =K v_R\EE
This scheme was then compared with lattice simulations in the Ising
limit that traditionally is considered a convenient laboratory to
study the properties of the theory. The result of this analysis
\cite{cea2003,cea2012} was $K=3.06   \pm 0.08 ~\rm{(stat)}  \pm 0.08 ~\rm{(syst)}$  or 
\BE M_h= 754 \pm 20 ~\rm{(stat)}  \pm 20 ~\rm{(syst)}  ~{\rm  GeV}
\EE 
The crucial question is then the following: is there any
experimental indication for $M_h$? Namely, if we identify $m_h=125 $
GeV, there could be one more massive excitation of the Higgs field
which fits with our $M_h$?

Here, we get in touch with a recent, independent analysis
\cite{cea2019} of the ATLAS + CMS data where experimental evidence
for an excess in the 4-lepton final state (at the 5$\sigma$ level)
was claimed. The natural interpretation of the excess would be in
terms of a scalar resonance around $700$ GeV which decays into two
$Z$ bosons and then into leptons. If the excess will be confirmed,
it could represent indeed the second heavier mass scale discussed in
this paper. This is not too far from the usual triviality bounds,
but the phenomenology of such heavy resonance (i.e. its production
cross sections and decay rates) may differ sizeably from the
perturbative expectations, see Ref.~\cite{journal}. For this reason,
we will stop here and wait for more experimental information, if
any.

\section*{Acknowledgements}

We would like to thank Paul Stevenson for information about Ref.~\cite{japan}
and for many useful discussions.


\begin{thebibliography} {99}
\bibitem{higgs}
P. W. Higgs, Phys. Lett. {\bf 12} (1964) 132.
\bibitem{englert}
F. Englert and R. Brout, Phys. Rev. Lett. {\bf 13} (1964) 321.
\bibitem{LHC}
ATLAS Collaboration, Phys. Lett. B {\bf 716} (2012) 1; CMS
Collaboration, Phys. Lett. B {\bf 716} (2012) 30.
\bibitem{lundow}
P. H. Lundow and K. Markstr\"om, Phys. Rev. E {\bf 80} (2009) 031104;
Nucl. Phys. B {\bf 845} [FS] (2011) 120.
\bibitem{japan}
S. Akyiama, et al. Phys. Rev. D {\bf 100} (2019) 054510.
\bibitem{mech}
M. Consoli and P. M. Stevenson, Int. J. Mod. Phys. A {\bf 15} (2000)
133.
\bibitem{stevenson2009}
P. M. Stevenson, Mod. Phys. Lett. A {\bf 24} (2009) 261.
\bibitem{CW}
S. Coleman and E. Weinberg, Phys. Rev. D {\bf 7}, (1973) 1888.
\bibitem{mpla1996}
M. Consoli and P. M. Stevenson, Mod. Phys. Lett. A {\bf 11} (1996)
2511.
\bibitem{gaussian}
T. Barnes and G. I. Ghandour, Phys. Rev. {\bf D22} (1980) 924; P. M.
Stevenson, Phys. Rev. {\bf D32} (1985) 1389.
\bibitem{zeit}
M. Consoli and P. M. Stevenson, Z. Phys. {\bf C63} (1994) 427; Phys.
Lett. {\bf B391} (1997) 144.
\bibitem{cea2003}
P. Cea, M. Consoli and L. Cosmai, Nucl. Phys. B (Proc. Suppl.) {\bf 129}
(2004) 780; hep-lat/0309050.
\bibitem{cea2012}
P.~Cea and L.~Cosmai,  ISRN High Energy Physics, vol. 2012, Article
ID 637950, arXiv:0911.5220.
\bibitem{cea2019}
P. Cea, Mod. Phys. Lett. A {\bf 34} (2019) 1950137.
\bibitem{further}
P. Cea, M. Consoli, L. Cosmai and P. M. Stevenson. Mod. Phys. Lett.
A {\bf 14} (1999) 1673.
\bibitem{stancu}
I. Stancu and P. M. Stevenson, Phys. Rev. D {\bf 42} (1990) 2710.
\bibitem{tedesco}
P. Cea and L. Tedesco, Phys. Rev. D {\bf 55} (1997) 4967.
\bibitem{journal}
P. Castorina, M. Consoli and D. Zappal\`a, J. Phys. G {\bf 35} (2008)
075010.
%
\end{thebibliography}
\end{document}